# Influence of Transmission Rank on EMF Exposure Measured With Provoked Data Traffic Around 5G Massive MIMO Base Stations

Lisa-Marie Schilling, Christian Bornkessel, Anna-Malin Schiffarth, Thanh Tam Julian Ta, Dirk Heberling, and Matthias Hein

*Abstract*— **The introduction of 5G New Radio networks with massive MIMO technology has complicated electromagnetic field exposure assessments for radiation protection. Massive MIMO transmission enables beamforming, beam steering, and spatial multiplexing across multiple transmission layers, with the number of simultaneous transmission paths depending on the rank of the radio channel, further named "transmission rank". Since the total transmission power of a base station is shared among these layers, rank variations affect the measured exposure levels, e.g., when assessments use provoked traffic via user equipment. This study investigates the impact of the transmission rank on the measured maximum exposure in the 3.6 GHz (n78) band of a German 5G network employing massive MIMO technology. Field measurements were performed using a spectrum analyzer with isotropic probe, to capture maximum field strengths under full-load traffic conditions. The transmission rank was manipulated by artificially degrading the reception quality of the user equipment with a shielding bag, forcing a single transmission layer (rank-1). The results were compared with unshielded operation allowing up to the maximum number of four independent transmission layers (rank-4). The data reveal exposure differences ranging from 1.7 dB to 5.4 dB, with a median of 4.3 dB at the measurement points studied. These findings highlight the necessity of considering the transmission rank in exposure assessments to electromagnetic fields.**

## I. INTRODUCTION

With the introduction of the 5G mobile radio network based on the 3GPP New Radio (NR) standard [1], data rates increase, interference, transmission and processing latencies as well as power consumption decrease, while the complexity of assessing the human exposure to electromagnetic fields (EMF) [2] increases significantly compared to previous mobile radio standards [3]. A key aspect of the introduction of 5G NR was the support of multiple-input-multiple-output (MIMO) systems with large adjustable antenna arrays, often referred to as massive MIMO systems. These systems enable beamforming and beam steering so that data can be simultaneously transmitted to spatially separated multiple user equipments (UE) logged into a cell. Spatial division multiple access (SDMA) allows different data streams to be transmitted over several physical layers, spatially separated via different beams or polarizations but with the same time-frequency resources [4,5].

Up to eight layers could be used according to the 5G NR standard [6]. However, current-state UEs are configured in such a way that they are limited to four layers [7]. The maximum possible number of layers of a data link corresponds to the transmission rank, which is determined by the characteristics of the radio channel, and indicates the number of possible simultaneous uncorrelated transmission paths. The transmission rank is reported by the rank indicator and is transmitted from the UE to the base station based on channel measurements [8]. If the rank is equal to one, referred to as "rank-1", only one layer can be used for transmission. In contrast, if channel conditions allow, up to four uncorrelated layers can be transmitted, referred to as "rank-4", thereby improving link quality and data throughput significantly. The total output power of the base station is then allocated to the transmission layers, for example it could be equally split [9].

There are currently two main methods for evaluating the maximum exposure of 5G massive MIMO base stations at a given measurement point (MP): Code-selective measurement of the signaling data contained in the cell-specific and load-independent synchronization signal and physical broadcast channel block (SSB), followed by subsequent extrapolation to maximum traffic load [3,10,11]. However, since SSB and user-centric data traffic are transmitted via different antenna patterns, a precise determination of the gain correction factor becomes challenging, especially when taking into account the MP-specific gain difference between traffic and broadcast beam in multipath-rich propagation channels [12]. Therefore, a frequency-selective measurement method has been established in which maximum traffic load is generated at the MP with the help of a UE [3,11]. However, as the UE can be served by the

The financial and technical support of this work by the Deutsche Telekom Technik GmbH is greatly acknowledged. This work has received funding by the German Federal Ministry of Research, Technology and Space (BMFTR) in the course of the 6GEM research hub under grant number 16KISK040. *(Corresponding author: Lisa-Marie Schilling.)*

Lisa-Marie Schilling, Christian Bornkessel, and Matthias Hein are with the Thuringian Center of Innovation in Mobility, RF and Microwave Research Group, Technische Universität Ilmenau, Germany (e-mail: lisa-marie.schilling@tu-ilmenau.de).

Anna-Malin Schiffarth and Thanh Tam Julian Ta are with the Institute of High Frequency Technology, RWTH Aachen University, Aachen, Germany.

Dirk Heberling is with the Institute of High Frequency Technology, RWTH Aachen University, Aachen, Germany and also with the Fraunhofer Institute for High Frequency Physics and Radar Techniques, Wachtberg, Germany.

base station via a location-dependent variable number of layers, this may lead to different exposure levels. Due to pathloss, shadowing, and interference specific for each individual transmission layer, the exposure is expected to depend on the number of layers. This paper investigates the impact of the transmission rank of a data link between base station and UE on the maximum exposure.

In Chapter II, the measurement methods and scenarios are described. The results of the exposure levels for various transmission ranks are presented and discussed in Chapter III. The conclusions from the results are provided in Chapter IV.

## II. MATERIALS AND METHODS

### A. Measurement Method

To assess the exposure levels for different transmission ranks, comparative measurements were taken at several measurement points around 5G massive MIMO base stations. The measurements were carried out in the n78 band at 5G NR [1] in the 3610–3700 MHz frequency band of the network operator Deutsche Telekom. The battery-powered portable spectrum analyzer SRM-3006 from Narda Safety Test Solutions with the associated isotropic field probe (type 3502/01, frequency range 0.42–6 GHz) in "Spectrum" mode was used as measurement device [13]. In this measurement mode, the spectrum is recorded with a selected resolution bandwidth, here RBW = 300 kHz, which is comparable to the frequency span of one resource block, and a video bandwidth of VBW = 5 kHz, to provide rms smoothing of the fluctuating signal envelope resulting from the digital modulation. In addition to displaying the current measured spectral field values, the maximum values occurring over a measurement period ("MaxHold" functionality) are also available. The measurement uncertainty was determined to be ±3 dB, corresponding to the expanded uncertainty for a coverage factor $k = 2$, i.e. 95% confidence interval [14].

The isotropic field probe was swept extensively with the measuring person turning around by 360° at each measurement point and sweeping the probe in the torso/head area at a height span of 0.8–1.8 m. This method is used to reliably record the spatial and temporal maximum exposure value, taking into account frequency-selective fading in the signal band. The maximum received field strength was evaluated from the spectrum at each measurement point, while the UE provoked full-load traffic conditions with the speed-test app FAST [15]. In view of the time-division duplex transmission mode, it was verified at each measurement point with the "Scope Mode" in time domain that the uplink exposure to the UE was below the downlink exposure of the 5G station.

A Samsung Galaxy S24 Ultra with the software QualiPoc [16] was used as the UE. This software enables to read out the number of layers, as illustrated by Fig. 1. With the help of a base station monitoring provided by the network operator, it was verified in preceding measurements that the value derived from QualiPoc agrees with the actual number of layers of the data link, and also that the app FAST retrieves the maximum resources of the mobile radio cell under investigation. The UE

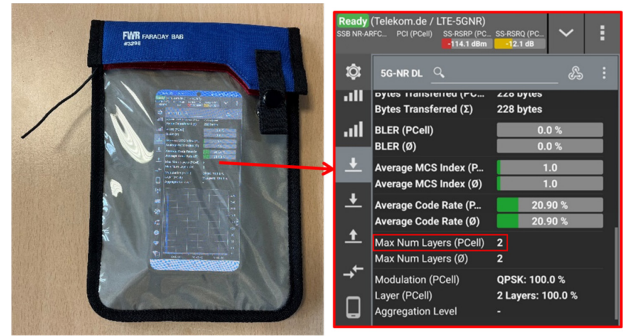

**Fig. 1.** UE in shielding bag with a wire fed through the closure (left) and exemplary UE monitoring screen to read out the number of layers ("Max Num Layers", right)

was forced to use the n78-band, whereby the LTE anchor band E-UTRA band 3 at 1800 MHz [1] also had to be switched on, in order to establish a data link in the non-stand-alone network.

In order to manipulate the transmission rank, a shielding bag [17] was used to artificially degrade the reception quality of the UE such as to enforce rank-1 or rank-2 transmission. As visualized in Fig. 1, a wire was fed through the closure of the shielding bag, in order to prevent the full shielding and maintain the reception quality of the mobile radio signals at a level that would allow the UE to connect to the base station.

During the measurement, the number of transmission layers was monitored via the QualiPoc display. At each measurement point, the measurements were performed with the UE shielded, to force a low rank, and without shielding, to enable transmission up to rank-4. Subsequently, the difference between these exposure results was evaluated. During measurements with and without shielding bag, the positions of the UE, the measurement person, and the sweeping volume were kept unchanged.

### B. Investigated Scenarios

Five measurement points were selected at each of two different 5G massive MIMO base stations, one from Huawei and one from Ericsson, with different Euclidean distances and line-of-sight (LOS) conditions between UE and base station antenna, as shown in Fig. 2 and listed in Table 1. In addition, the location of the measurement points varied in terms of their environmental conditions and thus also the multipath channel conditions. Some measurement points were located in a meadow, with propagation conditions similar to rural environment, while other measurement points fell into built-up sub-urban areas with multiple material interfaces provoking multipath propagation.

## III. RESULTS AND DISCUSSION

Figure 3 depicts the exposure differences for rank-1 and rank-4 transmission. At MP 1.3 and 1.4, rank-4 transmission was not possible due to the prevailing channel conditions, so that transmission with only three layers (rank-3) was occasionally detected during the measurement without shielding the UE.

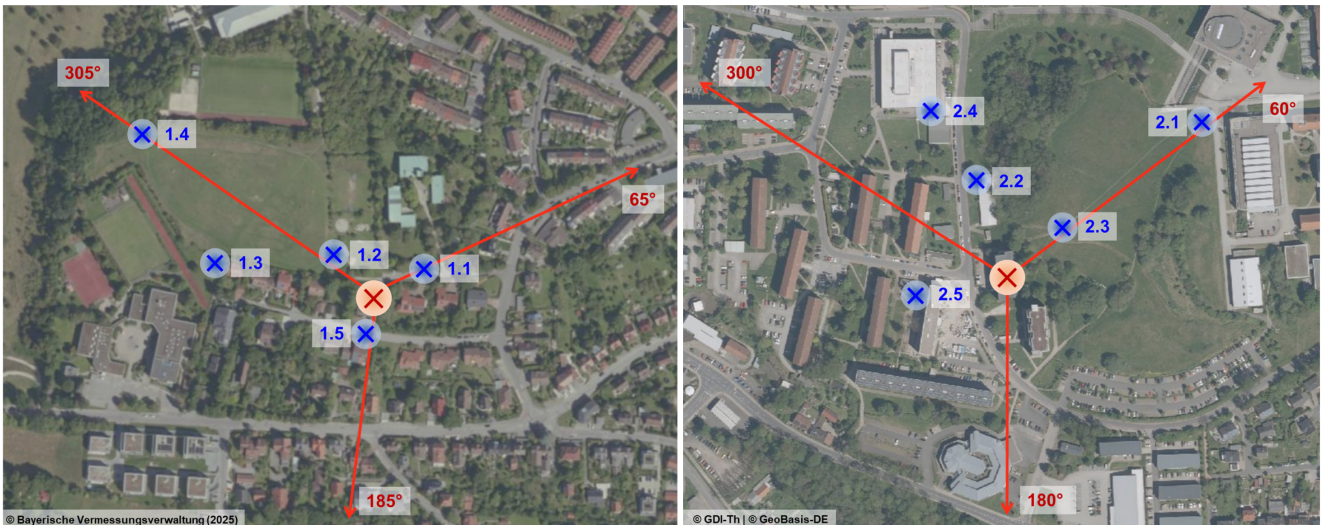

**Fig. 2.** Location of base station 1 (red) and measurement points MP 1.1–1.5 (blue) in Bavaria (left) and corresponding base station location 2 and MP 2.1–2.5 in Thuringia (right); the center direction of the three sectors are indicated by red arrows.

TABLE I
DISTANCE AND LINE-OF-SIGHT (LOS)-CONDITION OF
MEASUREMENT POINTS (MP) TO BASE STATION

| MP-No. | Distance to base station in m | LOS |
|---|---|---|
| **1.1** | 56 | No |
| **1.2** | 50 | Yes |
| **1.3** | 130 | No |
| **1.4** | 241 | No |
| **1.5** | 29 | Yes |
| **2.1** | 215 | Yes |
| **2.2** | 108 | No |
| **2.3** | 62 | Yes |
| **2.4** | 172 | No |
| **2.5** | 73 | No |

At all measurement points, the maximum exposure was higher for rank-1 than for rank-4 or rank-3. Overall, the difference in exposure levels across all measurement points varied between 1.7–5.4 dB, with the median being 4.3 dB. This effect exceeds the measurement uncertainty of ±3 dB significantly.

The resulting exposure difference between rank-4 and rank-1 transmission was observed to depend on the measurement point location. As up to two different beam directions with different polarizations are formed in a rank-4 transmission, it is reasonable to associate the rank value with the conditions of the respective propagation channels. Depending on the respective propagation scenarios, the exposure levels for different ranks can be lower or higher: If the measurement point were mainly served via a single dominant path (direction and polarization, rank-1), the exposure would decrease if power were shared with other layers (rank-4) that reach the measurement point with low intensity. If, on the other hand, there were several transmission paths of similar reception quality, the rank-1 exposure would supposedly remain equal or become lower in comparison to rank-4 transmission.

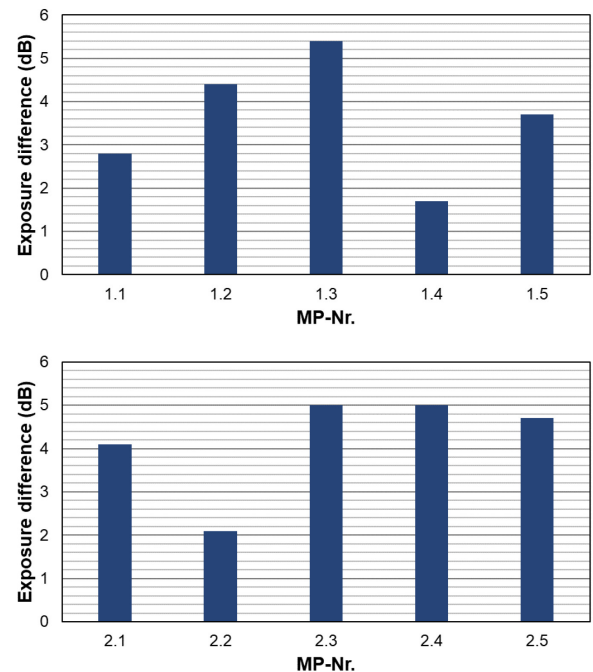

**Fig. 3.** Exposure difference between shielded and unshielded UE operation (low-rank – high-rank) at MP 1.1–1.5 (top) and MP 2.1–2.5 (bottom)

In the framework of our model, we observed one outlier in the measurement results at MP 1.4, where the smallest exposure difference between low-rank and high-rank was measured although this measurement point was located in an open meadow, where the line-of-sight would be expected to present the dominant transmission path. At this measurement point, we faced difficulties with staying connected to the 5G network at 3.6 GHz during the measurement run with the shielding bag to force rank-1 transmission. As a result, the sweeping process with the isotropic field probe could not be completed and thus may have corrupted the measurement.

## IV. Conclusion

The results of our study have proven that the number of transmission layers (rank) of 5G NR massive MIMO links has a significant influence on the maximum exposure levels when these are determined under provoked traffic conditions. Across all measurements, the exposure levels for low-rank (rank-1) and high-rank (rank-4, rank-3) transmissions differed, on average, by 4.3 dB, thus clearly exceeding the measurement uncertainty in most cases.

The exposure variations reflect the respective propagation channels involved in the communication links established. We are presently establishing a comprehensive data basis to enable systematic comparisons and conclusions.

These findings have direct implications for maximum exposure assessment in 5G massive MIMO networks based on provoked data traffic. These methods must consider the transmission rank, as neglecting may lead to significant underestimation of the maximum possible exposure.